\documentclass[aps,prl,amsmath,amssymb,amsfonts,showpacs,superscriptaddress,floatfix,twocolumn]{revtex4}

\usepackage{graphicx}
\usepackage{epsfig}
\usepackage[english]{babel}
\newcommand{\be}{\begin{equation}}
\newcommand{\bea}{\begin{eqnarray}}
\newcommand{\eea}{\end{eqnarray}}
\newcommand{\ee}{\end{equation}}

\def\p{{\cal P}}
\def\d{{\cal D}}
\def\one{\ensuremath{\hbox{$\mathrm I$\kern-.6em$\mathrm 1$}}}

\begin{document}

\title{Criticality, the area law, and the computational power of PEPS}

 \author{F. \surname{Verstraete}}
 \affiliation{Institute for Quantum Information, Caltech, Pasadena, US}
 \author{M.~M. \surname{Wolf}}
 \affiliation{Max-Planck-Institut f\"ur Quantenoptik, Hans-Kopfermann-Str. 1, Garching, D-85748, Germany}
 \author{D. \surname{Perez-Garcia}}
 \affiliation{Max-Planck-Institut f\"ur Quantenoptik, Hans-Kopfermann-Str. 1, Garching, D-85748, Germany}
\affiliation{\' Area de matem\' atica aplicada, URJC, M\' ostoles (Madrid), Spain }
 \author{J.~I. \surname{Cirac}}
 \affiliation{Max-Planck-Institut f\"ur Quantenoptik, Hans-Kopfermann-Str. 1, Garching, D-85748, Germany}

\begin{abstract}
The projected entangled pair state (PEPS) representation of
quantum states on two-dimensional lattices induces an entanglement
based hierarchy in state space. We show that the lowest levels of
this hierarchy exhibit an enormously rich structure including
states with critical and topological properties as well as
resonating valence bond states. We prove, in particular, that
coherent versions of thermal states of any local 2D classical spin
model correspond to such PEPS, which are in turn ground states of
local 2D quantum Hamiltonians. This correspondence maps thermal
onto quantum fluctuations, and it allows us to analytically
construct critical quantum models exhibiting a strict area law
scaling of the entanglement entropy in the face of power law
decaying correlations. Moreover, it enables us to show that there
exist PEPS within the same class as the cluster state, which can
serve as computational resources for the solution of NP-hard
problems.

\end{abstract}

\maketitle

The concept of entanglement plays a central role in both fields of
quantum information theory and of strongly correlated systems. In
\emph{quantum information theory} it lies at the heart of many
applications and it is viewed as a resource for various
information processing tasks. In \emph{condensed matter theory}
entanglement is one of the roots for the notorious complexity of
quantum many-body systems: its presence necessitates a description
within an exponentially growing Hilbert space and it is intimately
connected with many of the fascinating properties which quantum
matter can exhibit at small temperatures.

Many fundamental questions arise at the crossing of these fields:
how is entanglement related to the power of quantum computation on
the one hand, and the difficulties of classical simulations on the
other?  What is the scaling of the entanglement entropy in spin
systems, its relation to criticality, and the appearance of
topological quantum order? All these questions can be addressed
very easily within the framework of so--called projected entangled
pair states (PEPS)---this is the intention of this paper. We will
see in particular that all the above mentioned properties emerge
naturally already within the simplest classes of PEPS which
include cluster, toric code and resonating valence bond states.
This will enable us to settle a recent debate about the relation
between criticality and entropy scaling, and it allows us to find
computational resources for the solution of NP-hard problems. The
central tool of the paper is a general correspondence between
thermal states of classical 2D spin models and 2D quantum states
with a simple PEPS representation. This correspondence substitutes
thermal by quantum fluctuations while preserving the nature of
correlations, and it thus maps critical classical onto critical
quantum states.

We begin by recalling the PEPS formalism, which was introduced in
the context of numerical renormalization group methods for
simulating strongly correlated quantum spin systems
\cite{PEPS,VC04}. PEPS can be viewed as generalizations of the
AKLT valence bond solids \cite{AKLT} to arbitrary lattices and
dimensions. Consider an arbitrary connected graph where each of
$N$ vertices corresponds to a quantum system, a \emph{spin}, with
$d$ degrees of freedom. A PEPS $|\Psi\rangle\in \mathbb{C}^{d^N}$
is then constructed by (i) assigning to each vertex as many
virtual spins of dimension $D$ as there are adjacent edges, (ii)
putting a maximally entangled state $|I\rangle = \sum_{i=1}^D
|ii\rangle$ onto each edge, and (iii) mapping the virtual onto the
physical spins by applying a map
$P:\mathbb{C}^D\otimes...\otimes\mathbb{C}^D\rightarrow
\mathbb{C}^d$ at each vertex. Naturally, the graph is chosen
according to the physical symmetry, and although most of the
following holds in general we will consider square lattices
throughout.

The power of the PEPS formalism is based on two points. First,
every state has a PEPS representation \cite{Diego}. Hence, with
increasing $D$ this representation induces a hierarchy in the
space of states, from product states ($D=1$) to more and more
entangled ones. Second, it appears that many states arising in
physics are very well approximated by the lower levels of this
hierarchy \cite{approx}. This makes them a powerful variational
class for numerical renormalization group methods on the one hand
\cite{PEPS,Diego}, and an interesting testbed for all kinds of
quantum many-body questions on the other \cite{testbed}.

\emph{Quantum-classical correspondence:} Consider a classical
two-body spin Hamiltonian of the form
$H(\sigma_1,\cdots,\sigma_N)=\sum_{(i,j)}h(\sigma_i,\sigma_j)$
with $\sigma_i=1,\ldots,d$ and respective partition function
$Z=\sum_\sigma \exp[-\beta H(\sigma)]$ at inverse temperature
$\beta$. From this a corresponding quantum state can be
constructed by using the Boltzmann weights as superposition
coefficients such that \be |\psi_{H,\beta}\rangle=\frac1{\sqrt
Z}\sum_{\sigma_1,\cdots,\sigma_N} e^{-\frac\beta{2}
H(\sigma_1,\cdots,\sigma_N)}\; |\sigma_1,\cdots,\sigma_N\rangle\;.
\label{coh}\ee  We will see that $|\psi_{H,\beta}\rangle$ has the
following properties: (i) for diagonal observables it gives rise
to the same expectation values and correlations as the classical
thermal state, (ii) it has a simple PEPS representation with
 $D=d$, (iii) it is the ground state of a local quantum Hamiltonian, and (iv) when considering asymptotically
 large systems ($N\rightarrow\infty$) the scaling of the entropy of a block of spins obeys a strict area law.
Whereas (i) is a direct consequence of the construction, (iii) and (iv) are implied by the PEPS parametrization.
In order to see the latter we rewrite the state as \be\label{cohh} |\psi_{H,\beta}\rangle=\frac1{\sqrt
Z}\exp{\Big[-\frac\beta{2}\sum_{(i,j)}\hat{h}_{ij}\Big]}\;|+,\cdots ,+\rangle\;, \ee where
$|+\rangle=\sum_{s=1}^d |s \rangle$ and $\hat{h}_{ij}$ is a diagonal operator acting on sites $i,j$ as
$\hat{h}_{ij}\;|\sigma_i, \sigma_j\rangle = h(\sigma_i,\sigma_j)|\sigma_i, \sigma_j\rangle$. Following
Eq.(\ref{cohh}) we can think of the state $|\psi_{H,\beta}\rangle$ as being constructed from the product state
$|+,\cdots,+\rangle$ by applying (non-unitary) gates $\exp[-\beta \hat{h}/2]$ to all neighboring spins. In fact,
we may interpret Eq.(\ref{cohh}) as a quantum cellular automaton evolution in imaginary time. As explained in
\cite{VC04}, a nonlocal gate like $\exp[-\beta \hat{h}/2]$ can be reexpressed by local operations which act
additionally on an auxiliary maximally entangled state. More specifically, we take operators $\p$,
$\p':\mathbb{C}^{d^2}\rightarrow\mathbb{C}^{d}$, each acting as $\p|s,k\rangle =|s\rangle
\langle\varphi_s|k\rangle$. Then we obtain indeed $\exp[-\beta/2 \hat{h}]= \big(\p\otimes \p'\big)|I\rangle$ if
we choose the vectors $\varphi_s,\varphi'_{s'}$ such that $\sum_{k=1}^d \langle\varphi_s|k\rangle
\langle\varphi'_{s'}|k\rangle = h(s,s')$ which is always possible, e.g., by a singular value decomposition.
Applying these gates to all edges leads then to the desired PEPS representation.

As an example consider the ferromagnetic Ising model on a 2D
square lattice with
$$ H(\sigma)=-\sum_{(i,j)}\sigma_i\sigma_j\;,\qquad \sigma_i=\pm
1\;.$$ In this case we can choose $\varphi_s=\varphi'_{s}$ such
that $\langle\varphi_s|k\rangle$ are the matrix elements of the
square root of the matrix $h$. Applying all the gates gives then
rise to a PEPS \cite{GHZ} with
$$ P=|0\rangle\langle \varphi_0|\langle \varphi_0|\langle
\varphi_0|\langle \varphi_0|+|1\rangle\langle \varphi_1|\langle
\varphi_1|\langle \varphi_1|\langle \varphi_1|\;.$$ Clearly, the
expectation values of Pauli $S_z$ operators in
$|\psi_{H,\beta}\rangle$ equal the classical expectation values.
In the quantum case, however, we do not only have diagonal
observables, but also non-diagonal ones like $S_x$. Surprisingly,
their expectation values are determined by classical ones as well,
like
\begin{eqnarray}\nonumber
\langle\psi_\beta|S_x^1|\psi_\beta\rangle&=&\sum_{\sigma_1,\sigma_2,...}e^{-\frac\beta{2}\Big[H(\sigma_1,\sigma_2,\ldots)+H(-\sigma_1,\sigma_2,\ldots)\Big]}\\
&=& \sum_{\sigma_1,\sigma_2,...} e^{-\beta H(\sigma_1,\sigma_2,...)}\ e^{ \beta \sum_{(1,j)}
\sigma_1\sigma_j},\label{sx}
\end{eqnarray}
where the last term is a local 5-body expectation value in the
classical Gibbs state. In general, every local expectation value
in the quantum state corresponds to a local expectation value in
the classical state, where the region the observable acts on is
enlarged at most by the interacting neighborhood.

Before continuing it should be noted that several results related to the above classical-quantum correspondence
can be found in the literature: a connection between so--called Rokshar-Kivelson points and classical stochastic
models  was recently made in \cite{HAC} and between Hamiltonians and rapidly mixing reversible Markov chains in
\cite{Dorit}. In \cite{KZ} a generalization of the AKLT-state on 2D lattices was considered and demonstrated that
it can be mapped onto a classical vertex model. The PEPS formalism, discussed in the present paper, provides a
very natural framework for describing and generalizing those results.

\begin{figure}[t]
  \centering
\includegraphics[width=\linewidth]{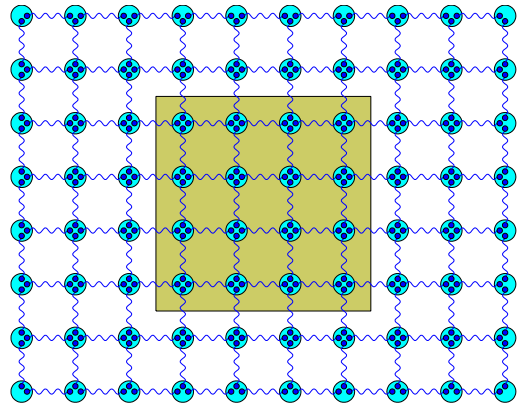}
  \caption{The entropy of a block of spins in a PEPS scales like the perimeter of the block: the Schmidt rank of the reduced density operator of the considered block of spins is bounded above by the product of the Schmidt ranks of the broken bonds.}
\end{figure}

\emph{Criticality and the area law:} Recently a lot of attention has been devoted to the scaling of the
entanglement entropy \cite{testbed,1D,Bosons,Fermions,poly}. That is, given a ground state, how does the entropy
of a contiguous subsystem scale with the size of the latter? Originally appearing in the context of black holes
the renewed interest in this question comes from the investigation of quantum phase transitions and the quest for
powerful ansatz-states for the classical simulation of quantum systems. In 1D it is known that critical states
corresponding to a conformal field theory exhibit a logarithmic divergence of the entropy $S\sim\log L$ ($L$
being the length of the subsystem), whereas there seems to be a saturation for all non-critical systems
\cite{1D}. In $\d>1$ dimensions quasi-free systems of Bosons \cite{Bosons} and Fermions \cite{Fermions} have been
studied. Whereas in the non-critical (gapped) bosonic case there is a strict area law $S=O(L^{\d-1})$ (for a cube
with edge-length $L$), this is violated in the gapless case of Fermions, where $S\sim L^{\d-1}\log L$. This
naturally rises the question about a one-to-one correspondence between criticality and a violation of the area
law. The PEPS formalism together with the above classical-quantum correspondence enables us now to answer this
question in the negative in a very simple way: consider the classical 2D-Ising system, which is known to become
critical in the thermodynamic limit at $\beta_c=\frac12\big(1+\sqrt{2}\big)$. The corresponding quantum state
$|\psi_{\beta_c}\rangle$ will then have exactly the same correlations $\langle S_z^0S_z^r\rangle \sim
1/\sqrt{r}$, which now reflect critical quantum rather than thermal fluctuations. In spite of this, the state
obeys a strict area law bound $S\leq 4 L$ due to the PEPS representation with $D=2$ since the entropy is solely
generated by breaking $4L$ entangled bonds at the boundary (see figure 1). In the following we will show that,
unlike the cases in \cite{poly}, the power law decay of correlations is not a consequence of long range
interactions. Rather $|\psi_{\beta_c}\rangle$ is the ground state of a \emph{local} Hamiltonian.

\emph{Parent Hamiltonians: } Every PEPS with finite $D$ is (on a sufficiently large lattice) the ground state of
a local Hamiltonian. The standard construction of such \emph{parent Hamiltonians} identifies projectors onto null
spaces of reduced density operators with the interaction terms in the Hamiltonian \cite{AKLT,testbed}. Here we
will follow a  different approach, related to the one described in \cite{Dorit,HAC}, which is adapted to PEPS
constructed from classical models and allows us to prove uniqueness and existence of a gap above the ground state
energy for $\beta<\beta_c$.

Consider any ergodic local Markov process obeying detailed balance
and converging to the equilibrium distribution of the classical
 model at inverse temperature $\beta$ (e.g., by using Metropolis Monte Carlo
 or general Glauber dynamics \cite{Landau}).
The stochastic transition matrix corresponding to the Markov
process can be written as a sum $M(\beta)=\sum_{k} M_k(\beta)$
where each $M_k(\beta)$
 acts locally and obeys detailed balance. The latter requires
\[p_a(\beta) \big[M_k(\beta)\big]_{a,b}=p_b(\beta) \big[M_k(\beta)\big]_{b,a}\]
with $p_a(\beta)= \exp(-\beta H(a))/Z$ and $a,b$ denoting a
particular configuration of the $N$ spins. In matrix notation,
this is equivalent to imposing that all the matrices
\[P_k(\beta)=e^{-\frac{\beta}{2} \sum_{(ij)}\hat{h}_{ij}}\ M_k(\beta)\ e^{\frac{\beta}{2} \sum_{(ij)}\hat{h}_{ij}}\]
are symmetric. Obviously, the operator $\sum_k P_k(\beta)$ is
symmetric and has exactly the same eigenvalues as $\sum_k
M_k(\beta)$ since they are connected by a similarity
transformation. Furthermore, all $P_k(\beta)$ are local operators
if the $M_k(\beta)$ were local. Note that $\openone-\sum_k
M_k(\beta)$ only has non-negative eigenvalues with the equilibrium
distribution corresponding to eigenvalue $0$. We thus define the
Hamiltonian $H(\beta)=\openone-\sum_k P_k(\beta)\geq 0$ which is a
sum of local operators such that by construction
$|\psi_\beta\rangle$ is the ground state of $H(\beta)$. Moreover,
$H(\beta)$ is gapped iff the stochastic matrix $M(\beta)$ has a
gap. In this way the gap in the quantum Hamiltonian corresponds to
the rate of convergence to equilibrium of the Markov process. In
fact, the existence of a gap in $M(\beta)$ for $\beta<\beta_c$ was
proven in \cite{Martinelli} for a class of models including the 2D
Ising model. At precisely the critical point, Monte Carlo methods
exhibit a slowing down, leading to a gapless critical quantum
Hamiltonian. In fact, power law decaying correlations imply that
the Hamiltonian has to be gapless \cite{Hgap}.

\emph{Computational power of PEPS: } In this section we will treat the PEPS as a resource for computational
tasks. Given a source which produces a specific state in an efficient manner, together with the ability of
performing arbitrary local measurements, what kind of computational problems can we solve efficiently? This
question is clearly inspired by the \emph{cluster state} computational model \cite{Briegel}. In fact, it was
shown in \cite{VC04} that the cluster state is a PEPS with $D=2$. Moreover, it is known to be a resource state
for universal quantum computation, i.e, it enables us for instance to solve a typical NP
problem---factorization---by merely performing local measurements. Since PEPS with $D=1$ are product states,
$D=2$ is in fact the simplest class in which useful resources  can be expected. Exploiting the above formalism it
is now simple to show that there are other powerful resource states within this class, which even enable the
efficient solution of NP-hard problems. In order to see this, note that given a quantum state
$|\psi_\beta\rangle$ which corresponds to a classical Hamiltonian $H(\sigma)$ we can efficiently determine
expectation values in the classical Gibbs state by performing local (diagonal) measurements on
$|\psi_\beta\rangle$. For $\beta\rightarrow\infty$ we can for instance \emph{measure} the ground state energy
\cite{advantage}. This is in particular true for the $D=2$ PEPS corresponding to a two-dimensional Ising spin
glass within a magnetic field. This task was, however, shown to be an NP-hard problem in \cite{Barahona} together
with the case of 3D spin glasses without magnetic field. Similarly,  the determination of the partition function
of the Potts model ($D>2$) is known to be $\#P$-hard and tightly connected to hard problems in knot theory. As
the task of calculating expectation values of PEPS can be done by contracting a network of tensors arranged on a
square lattice \cite{PEPS}, the above arguments prove that such a contraction of tensors is in general a NP-hard
problem in the number of tensors \cite{n20}.

The approach of encoding the solution to an NP-hard problem into a quantum state is reminiscent of adiabatic
quantum computing \cite{adiabatic} which, however, deals typically with ground states of 1D albeit non-local
Hamiltonians. In fact, one possible way of generating PEPS would be by adiabatic means with the usual caveat
concerning the gap of the system. However, as in the case of the cluster state, there might be better ways of
generating these states since after all we have an efficient \emph{local} parametrization---the above observation
makes the generation of PEPS a highly interesting problem.

Let us finally show that two other classes of states, important
for quantum information and condensed matter theory, are contained
within small-$D$ PEPS as well:

\emph{Toric code states } introduced in the context of quantum
error correction are very interesting as they exhibit nontrivial
topological properties \cite{Kitaev}. In the case of an infinite
square lattice, the toric code is the ground state of a
Hamiltonian consisting of local commuting projectors, each of them
annihilating the ground state. The state can again be written in
terms of the (zero-temperature) Boltzmann weights of a classical
statistical model
\[|\psi_{tor}\rangle\simeq
\lim_{\beta\rightarrow\infty}\exp\left(+\frac{\beta}{2}\sum_{\square_i}S_z^{\alpha_i}S_z^{\beta_i}S_z^{\gamma_i}S_z^{\delta_i}\right)|++\cdots
+\rangle\] where $\square_i$ denotes the i'th plaquette in the lattice with the spins on the edges. We can again
represent the commuting nonlocal gates by introducing entangled auxiliary degrees of freedom and applying local
operations. More specifically, the 4-qubit gate can be implemented by distributing the states $|I\rangle$ between
the 4 qubits, followed by local projections of the form $|0\rangle\langle\Psi^+|+|1\rangle\langle\Psi^-|$ where
$|\Psi^\pm\rangle=|00\rangle\pm|11\rangle$.


Implementing this map on all plaquettes of the lattice, we obtain
projectors of the form
\begin{eqnarray*}
P_e&=&|0\rangle\langle\Psi^+|_{12}\langle\Psi^+|_{34}+|1\rangle\langle\Psi^-|_{12}\langle\Psi^-|_{34}\\
P_o&=&|0\rangle\langle\Psi^+|_{14}\langle\Psi^+|_{23}+|1\rangle\langle\Psi^-|_{14}\langle\Psi^-|_{23}
\end{eqnarray*}
where $P_e$, $P_o$  act on the even and odd sites of the bipartite
lattice respectively (the labels 1..4 denote the virtual qubits in
clockwise order). Hence it is again a simple PEPS with $D=2$
exhibiting nontrivial topological behavior. In the case of PEPS
with finite $D$, it is indeed always simple to calculate the
topological entropy defined in \cite{PK} explicitly---the PEPS
formalism seems to provide a
 promising avenue for generating other states exhibiting those
fascinating properties.

\emph{Resonating valence bond states} (RVB) have been studied extensively in the context of strongly correlated
systems \cite{Anderson}. These states exhibit topological quantum order and do not seem to have any classical
statistical model associated to them because the wave function contains negative weights. For the case of
simplicity, let us consider the simplest RVB state which is the equal weight superposition of all possible
coverings of singlets over nearest neighbors on a square lattice.
It can easily be checked that this RVB is equivalent to the PEPS defined by \bea P &=&|0\rangle\left(\langle
0222|+\langle 2022|+\langle 2202|+\langle 2220|\right)\nonumber\\
&&+ |1\rangle\left(\langle 1222|+\langle 2122|+\langle 2212|+\langle 2221|\right)\nonumber\eea acting on virtual
\emph{singlets} of the form $|S\rangle=|01\rangle-|10\rangle+|22\rangle$ distributed between all nearest
neighbors. Interestingly, we need $D=3$ in this case, and again the area law is automatically proven (i.e., the
entropy of a block of spins scales like the boundary). In a similar way, RVB with singlets distributed beyond
nearest neighbors can easily be constructed.

\emph{In summary} we found that already the lowest levels of the
PEPS hierarchy exhibit an enormously rich structure---they contain
highly interesting states for quantum information (e.g., cluster
and toric code states) as well as for condensed matter theory
(e.g., critical and RVB states). This makes them an interesting
variational class for numerical methods and a rich testbed for
quantum many-body questions. Based on a classical-quantum
correspondence we were able to find critical quantum models whose
entropy scaling contrasts with the one for Fermions (and
corresponding spin models \cite{JW}). Moreover, it yielded a local
description of simple PEPS encoding the solution of NP-hard
problems.

\acknowledgements Work supported by the Gordon and Betty Moore Foundation,  European projects, der Bayerischen
Staatsregierung and MEC MTM-2005.


\begin{thebibliography}{99}
\bibitem{advantage} One could argue that this holds as well for the
classical ground state. The point is, however, that the latter is
merely characterized as a solution of a global optimization
problem, whereas the PEPS is explicitely given by local operators
$P$. Moreover, due to the separation of energy levels  the
statistics of the measurements is not an issue (similar to
\cite{adiabatic}).
\bibitem{n20} Note that this does not compromise the validity of the
simulation schemes based on PEPS developed in \cite{PEPS}, as the problem of $NP$-hardness does not apply in the
case of e.g. translational invariance.
\end{thebibliography}
\end{document}